\documentstyle[version2,preprint,aps]{revtex}
\begin{document}
\draft

\rightline{cond-mat/9406002}

\begin{title}
Measuring the Relative Phase of the Energy Gap

 in a
High-Temperature Superconductor

 with EELS
\end{title}

\author{Michael E. Flatt\'e}

\begin{instit}
Division of Applied Sciences, Harvard University,
Cambridge, Massachusetts 02138
\end{instit}

\begin{abstract}
A method of measuring the relative phase of the energy gap
in a high-temperature superconductor is suggested for electron
energy loss spectroscopy. Energy-resolved measurements of
off-specular scattering should show a feature  similar to
the specular feature associated with the gap. Unlike the
specular feature, which reflects an average of the gap over
the (normal) Fermi surface, the energy loss of the
off-specular feature depends on the
superconducting energy gap at only two locations on the Fermi
surface. The onset of the feature reflects the relative
phase between these two points. This result is independent of
surface characteristics. Such characteristics affect the {\it magnitude} of the
off-specular feature, not its location or onset. The size of the feature
is estimated for a simple surface model.  Implications of
specific measurements on $\rm Bi_2Sr_2CaCu_2O_8$ are
discussed.
\end{abstract}
\vfill\eject

\narrowtext

The details of the energy-gap anisotropy in the high-temperature
superconductors remain a critical unresolved issue.  Despite recent
advances in experiment, yielding strong evidence for a finite density
of states at low energy\cite{Hardy} and angular
anisotropy in the gap\cite{Shen,Wollman}, whether the gap has nodes or
changes sign is not known.

This Communication will suggest that the gap magnitude and
relative phase around the Fermi surface may be measured with
electron-energy-loss-spectroscopy (EELS). The
proposal is related to a method proposed earlier
using inelastic
neutron scattering measurements of phonon linewidths\cite{Flatte,FQS}.
The angle-resolved influence of the gap on the phonon linewidth
occurs due to the single-electron-single-hole decay channel of the
phonon.
The neutron-scattering experiment requires accurate
measurements of the difference in phonon linewidths between the
normal and the superconducting state. The small normal-state
electronic linewidths in the high-temperature superconductors $\rm
La_{1.85}Sr_{0.15}CuO_4$\cite{lasr}, $\rm
YBa_2Cu_3O_7$\cite{Pintschovius}, and
$\rm Bi_2Sr_2CaCu_2O_8$\cite{Mook} render these measurements
quite difficult.

Angle-resolved off-specular electron-energy-loss spectra are sensitive
to single-electron-hole-pair scattering processes, and thus
to the angle-resolved gap magnitude and phase. On-specular
EELS has been successfully performed on
$\rm Bi_2Sr_2CaCu_2O_8$\cite{Lieber}, but these
near-zero-momentum scattering processes yield features which
depend on an average of the gap. Sharp features should
appear in the loss spectra for momentum transfers greater
than an inverse superconducting coherence length even if the
gap is anisotropic. These features depend on the
angle-resolved gap magnitude and phase at points on the Fermi surface
which differ for each momentum transfer.

Simply from kinematics, an
electron scattering by parallel crystal
momentum ${\bf Q}$ with energy loss $\omega$ in a single scattering event
can only\cite{waffle} be
affected by the superconducting
electrons through the imaginary part of the
polarization,  ${\rm Im}P^{ret}_{bulk}({\bf Q},\omega)$.
The subscript {\it bulk}
indicates the polarization is calculated for an infinite,
nonterminated crystal. The surface normal is arranged parallel to the
trivial electronic direction $(\hat z)$.
Due to the
two-dimensional electronic structure of the superconductor,
no momentum variable in the
$\hat z$ direction is required for
$P^{ret}_{bulk}({\bf Q},\omega)$.
The details
of the surface will affect the magnitude of
the feature in the scattering probability which originates
from $P^{ret}_{bulk}({\bf Q},\omega)$.

Now the structure of $P^{ret}_{bulk}({\bf Q},\omega)$ for $Q>2\pi\xi^{-1}$,
where $\xi$ is the superconducting coherence length, will be
discussed. The discussion is more detailed in Refs.\cite{Flatte,FQS}.
Only the first-order polarization diagram produces sharp features,
so only that diagram, a two-quasiparticle bubble, will be considered
here.
In the superconductor there is a minimum energy to
create two electronic quasiparticles due to the gap. In an isotropic
superconductor that minimum energy is just twice the energy gap. For
a loss
$\omega$ less than twice the gap $P^{ret}_{bulk}({\bf Q},\omega)$
is purely real; there is no
scattering with loss $\omega$
due to the superconducting electrons. For
$\omega$ greater than twice the gap there are electronic scattering
processes. The transition between the two in ${\rm
Im}P^{ret}_{bulk}({\bf Q},\omega)$ is discontinuous, as
predicted\cite{Bobetic,Schuster} and seen\cite{Shir3,Shirane} in the
low-temperature superconductors $\rm Nb_3Sn$ and niobium in
phonon linewidths (which also measure ${\rm Im}P^{ret}_{bulk}({\bf
Q},\omega)$).

In a material with a
quasi-two-dimensional Fermi surface, the two quasiparticles can only
be created at certain locations on the Fermi surface, as shown in
Fig.~1 (the $\rm Bi_2Sr_2CaCu_2O_8$ Fermi surface\cite{Dessau}).
Both quasiparticles of a pair must be created
very near the Fermi
surface. This is due to the large electronic energy scale of the
Fermi surface (eV) compared with the
losses under consideration of the scattering electrons (meV).
Thus the minimum loss $\tilde\Delta({\bf Q})$ for an electronic
scattering event with momentum transfer ${\bf Q}$ would
depend on the sum of the gap
magnitudes at the two points where quasiparticles are
created. Those points are determined by placing the momentum
transfer
on the Fermi sea such that both the head and the tail lie on the
Fermi surface.  The quasiparticles are created at the head of the
vector and the Fermi surface point opposite the tail.
$\tilde\Delta({\bf Q})$ is frequently multivalued. As shown in
Fig.~1, for a momentum $0.5(\pi,\pi)$, there
are three possible pairs of quasiparticles (labeled $1-3$ on Fig.~1).
Each of these three thresholds may be independently observed as a
discontinuity in the differential scattering of electrons.

$Q$ must be greater than $2\pi\xi^{-1}$ or the quasiparticles can be
created anywhere on the Fermi surface.  EELS experiments which
detect the gap in high-temperature superconductors have so far
only been reported for near-zero momentum transfer\cite{Lieber}.

The polarization $P^{ret}_{bulk}({\bf Q},\omega)$
in the superconducting state arises from two
contributions. There is a ``normal'' polarization, where the vertex
creates an electron and hole, which recombine at the other vertex.
There is also an ``anomalous'' polarization, where the vertex creates an
electron and hole, the electron turns into a hole through interaction
with the gap, the hole turns into an electron, and then they recombine
at the other vertex. The phase of the gap does not affect the normal
process, but the anomalous process depends on the relative phase
$\Delta\phi$ of the gap at the two points where quasiparticles are
created. The sum of these two processes, at the threshold for
electronic scattering processes, determines whether the discontinuity
is its full size relative to the normal-state scattering probability at
$T_c$. The ratio of the discontinuity $\Delta (d^2S/d\Omega d\omega)$
to the normal-state differential probability at $T_c$,
$d^2S_N/d\Omega d\omega$, is
\begin{equation} {\Delta (d^2S/d\Omega d\omega)\over
d^2S_N/d\Omega d\omega} = {\pi\over 2}\cos^2\left({\Delta\phi\over
2}\right).
\end{equation}
The full energy dependence at $T=0$ is
\begin{eqnarray}
{d^2S(\omega)/d\Omega d\omega\over
d^2S_N(\omega)/d\Omega d\omega} =
\theta(\omega-&&\tilde\Delta({\bf Q}))\Bigg\{
\cos^2\left({\Delta\phi\over 2}\right)E\left(\left[1-\left({\tilde\Delta({\bf
Q})\over \omega}\right)^2\right]^{1/2}\right)\nonumber\\
+\sin^2\left({\Delta\phi\over
2}\right)&&\left[\left({\omega+\tilde\Delta({\bf Q})\over
\omega}\right)E
\left({\omega-\tilde\Delta({\bf Q})\over
\omega+\tilde\Delta({\bf Q})}\right)
- {2\tilde\Delta({\bf
Q})\over\omega}K \left({\omega-\tilde\Delta({\bf Q})\over
\omega+\tilde\Delta({\bf Q})}\right) \right]\Bigg\}\label{fulle}
\end{eqnarray}
where $E$ and $K$ are complete elliptic functions and
$\theta$ is the Heavyside step function.

The normalized differential probability (Eq. (2)) is plotted in
Fig. 2 for three values of
$\Delta\phi$. In the isotropic superconductor,
$\Delta\phi=0$ for all momenta transfers. For a
$d_{x^2-y^2}$ gap\cite{DJS,DP}, which is real but changes sign,
$\Delta\phi$ can be $\pi$. An intermediate value of
$\Delta\phi$ can only occur for a complex gap, which breaks
time-reversal symmetry.
Since the EELS
experiments can be performed at very low substrate
temperatures, the calculations presented here  are for $T=0$.

Ref.\cite{FQS} contains examples of the
detailed temperature dependence of $(d^2S(T)/d\Omega
d\omega)[d^2S_N/d\Omega d\omega]^{-1}$.
This is the same quantity which
is plotted there as the ratio of phonon linewidths at $T$ and
$T_c$, ($\gamma(T)/\gamma(T_c))$.
The above arguments concerning the
structure of ${\rm Im}P^{ret}_{bulk}({\bf Q},\omega)$
are independent of the surface
model which is now introduced to estimate the magnitude of
$(d^2S(T_c)/d\Omega d\omega)$.

The scattering by superconducting electrons is through the polarization
diagram --- therefore the equations from dipole scattering theory\cite{Mills}
can be used.
{}From Eq. (3.20) of Ref. \cite{Mills},
the differential probability for scattering
into final angle $\Omega$ with energy loss $\hbar\omega$
is
\begin{equation}
{d^2S\over d\Omega d\omega} = {2e^2m^2|R|^2\over
\pi\cos\theta_s}\left({ k_S\over k_I} \right){P({\bf
Q},\omega)\over [Q^2+(k^z_I-k^z_S)^2]^2}\label{dsdo}
\end{equation}
where $R$ is the reflection coefficient of the incident and
scattered beam, $k_S$ is the momentum and $k^z_S$ is the
$z-$momentum of the scattered beam,
$k^z_I$ is the $z-$component of the momentum of the incident
beam, $k_s^z = k_s\cos\theta_s$,
$e$ and $m$ are the charge and mass of
the electron, and
\begin{equation}
P({\bf Q},\omega) = \int {\rm e}^{-i\omega t}
dt\int_{-\infty}^0 dz\int_{-\infty}^0 dz'
\exp[Q(z+z')]\langle\rho({\bf Q},z',t)\rho(-{\bf
Q},z,0)\rangle. \label{psurface}
\end{equation}
The small-angle condition is
used twice to derive
Eq. (\ref{dsdo}): (a) processes where the scattering
interaction provides the momentum to reverse the sign of the
electron's perpendicular momentum are ignored and (b) the
electron beam is assumed not to penetrate substantially into
the crystal compared to the range of influence of the
density fluctuations on the electric field outside the crystal.
Treating (a) correctly would
provide a correction of order $Q^2/[(k_I^z-k_S^z)^2+Q^2]$. For
the incident energies (20eV) of interest, and considering the
smallest momentum transfer of interest ($\sim 0.4\AA^{-1}$), this
correction is of order $2\%$. Zone-boundary-momentum-transfer
scattering, however, may have a substantial contribution from these
corrections. (b) is not a concern because photoemission
experiments\cite{Photo} indicate an energetic electron could not
penetrate more than one unit cell without being multiply scattered.
Both assumptions (a) and (b) lead to an underestimate of the size of
the effect. Certain effects which are difficult to model
and which will decrease the actual size of the effect, such
as multiple scattering and scattering from other excitations, will not
be considered in this estimate.

We now discuss the structure of $P({\bf Q},\omega)$.
In a material with a quasi-two-dimensional electronic structure,
$P^{ret}_{bulk}({\bf q},\omega)$ is independent of $q_z$. For simplicity the
surface is assumed to be a hard wall potential for the superconductor's
electrons.  Then $P({\bf Q},\omega)$ can be
expressed in terms of $P^{ret}_{bulk}({\bf Q},\omega)$ as
\begin{equation}
P({\bf Q},\omega) = {2e^2\over \pi Q}
\left[{\rm tan}^{-1}\left({Q\over q_m}\right) - {q_m Q \over
Q^2+q_m^2}\right]{\rm Im}P^{ret}_{bulk}({\bf Q},\omega),
\label{pstopb}
\end{equation}
where $q_m$ is the zone-boundary momentum in the $z-$direction.
The assumptions made
about the surface influence the size of the effect through Eq.
(\ref{pstopb}), but the proportionality to $P^{ret}_{bulk}({\bf Q},\omega)$
is independent of the surface model. The final form of Eq.
(\ref{dsdo}) is
\begin{equation}
{d^2S\over d\Omega d\omega} = {4e^4m^2|R|^2\over
\pi^2Q\cos\theta_s}\left({k_S\over
k_I}\right)\left({\rm tan}^{-1}\left({Q\over q_m}\right) - {q_mQ\over
Q^2+q_m^2}\right)
{{\rm Im}P^{ret}_{bulk}({\bf Q},\omega)\over
[Q^2+(k_I^z-k_S^z)^2]^2}.\label{dsdof} \end{equation}
For a momentum transfer of the inverse coherence length, and
using the density of states information from
photoemission\cite{Dessau}, $5$ states/(eV Cu site),
the magnitude of the effect is $\sim 1/$(steradian
eV).

This Communication concludes with the identification of specific momenta
transfers of interest in $\rm Bi_2Sr_2CaCu_2O_8$. The Fermi
surface is shown in Fig.~1\cite{Dessau}. The momentum transfer
$0.5(\pi,\pi)$, whose quasiparticle channels are shown as vectors
$1-3$, corresponds to the momentum of a phonon whose linewidth was
tentatively identified as narrowing in Ref. \cite{Mook}. The points on
the Fermi surface probed by this momentum transfer are shown in
one-eighth of the zone by the labeled filled triangles. Each of these
thresholds may be independently observed if the gap is anisotropic.
If the gap has $d_{x^2-y^2}$ symmetry\cite{DJS,DP},
the momentum transfer $q_n$ should have an extremely low threshold
loss.
An additional check for $d_{x^2-y^2}$ symmetry is that the
onset of a gap feature for all scattering momenta parallel
to the $\Gamma-M$ direction should be sharp ($\Delta\phi=0$). The
onset for scattering momenta parallel to $\Gamma-X$ should be
smooth and weak ($\Delta\phi=\pi$).

I wish to acknowledge several illuminating discussions with C.M.
Lieber and R.B. Phelps and the support of the U.S. Joint
Services Electronics Program (JSEP) through ONR
N00014-89-J-1023.

\figure{Vectors $1-3$ are the non-degenerate ways to
position the momentum $0.5(\pi,\pi)$ on the $\rm
Bi_2Sr_2CaCu_2O_8$ Fermi surface (dashed line)\cite{Dessau}.
Observation of off-specular gap features with this momentum
transfer would probe the gap at the points $1-3$ indicated
by filled triangles. The onset of such features would need to
be smooth and weak to be consistent with a $d_{x^2-y^2}$ gap
symmetry. Any gap feature with momentum transfer $\vec{q_n}$ must
occur at very low-frequency to be consistent with
$d_{x^2-y^2}$. An isotropic gap would have gap features in
the scattering probability at energies similar to the
on-specular feature\cite{Lieber}.}

\figure{Scattering probability per unit energy and
angle at $T=0$, normalized to the
normal-state scattering probability. Solid line: relative phase
$\Delta\phi=0$ between  gaps at two places quasiparticles are created
on the Fermi surface. Short dashed line: relative phase
$\Delta\phi=\pi$. Long dashed line: relative phase $\Delta\phi=\pi/2$.}


\begin{references}
\bibitem{waffle} Surface reconstruction and spin-flip scattering are
ignored.
\end{references}
\end{document}